\documentclass[conference]{IEEEtran}
\usepackage{ amssymb }
\usepackage{color}
\usepackage{mathtools}

\DeclareMathOperator*{\argmin}{arg\,min}

\usepackage{amsthm}
\usepackage{amsmath}
\newtheorem{theorem}{Theorem}

\newtheorem{proposition}[theorem]{Proposition}

\title{$K$ Users Caching Two Files: An Improved Achievable Rate}

\author{\IEEEauthorblockN{Saeid Sahraei\IEEEauthorrefmark{1},
Michael Gastpar\IEEEauthorrefmark{2}}
\IEEEauthorblockA{School of Computer and Communication Sciences,
EPFL\\
Lausanne, Switzerland\\
Email: \IEEEauthorrefmark{1}saeid.sahraei@epfl.ch,
\IEEEauthorrefmark{2}michael.gastpar@epfl.ch}}

\begin{document}
\maketitle
\begin{abstract}
Caching is an approach to smoothen the variability of traffic over time. Recently it has been proved that the local memories at the users can be exploited for reducing the peak traffic in a much more efficient way than previously believed. In this work we improve upon the existing results and introduce a novel caching strategy that takes advantage of simultaneous coded placement and coded delivery in order to decrease the worst case achievable rate with $2$ files and $K$ users. We will show that for any cache size $\frac{1}{K}<M<1$ our scheme outperforms the state of the art. 
\end{abstract}

The performance of content delivery services is highly dependent on the habits of the users and how well the servers model these habits and adapt their content distribution strategies to them. A basic observation of these habits is the temporal variability of the demands which in its simplest form can be formulated as high congestion during a particular time interval and low traffic for the rest of the day. One popular mechanism that the network can adapt to cope with this issue is caching: during the low traffic time interval, typically mornings, the servers store parts of the content in local memories of the users which may be helpful in the evenings, and hence reduce the peak traffic load. A notable challenge with this strategy is that typically the servers are not aware of which contents will be requested by the users in the peak time. Therefore, the caching of contents in local memories must be performed in such a way that regardless of what requests the users make, the contents are still helpful in reducing the traffic, as much as possible.

Perhaps the simplest solution to this problem is to partially store every file at the local caches of the users and transfer the rest of the data uncoded according to the demands made in the delivery time. In their seminal works  \cite{maddah2014fundamental,maddah2013decentralized} Maddah-Ali and Niesen have proved that by using network coding techniques this simple strategy can be significantly outperformed if one allows coding across different files on the server and jointly optimizes the caching and the delivery strategies.

Despite its impressive potentials, the caching strategy introduced in \cite{maddah2014fundamental} is known to perform poorly when the cache size is small, and in particular when the number of users is much larger than the number of files, $K\gg N$. The applicability of this paradigm in real world scenarios is manifold. A good example is when the files on the server vary widely in their popularity. It has been proved \cite{niesen2014coded} that a nearly optimal caching strategy is to group files with similar popularities and ignore caching opportunities among files from different groups. The cache of each user is then divided into several segments, each segment dedicated to one such group. If the number of groups is large, then the cache size dedicated to each group, as well as the number of files within each group will be small. Another case appears when there are few popular television hits, say on Netflix, which are 
streamed by millions of users across the world.

In this work we take a step in improving the performance of caching strategies for small cache size $M$  when $K\gg N$. This scenario has been studied before \cite{chen2014fundamental} where a new point $(M,R)=(\frac{1}{K},\frac{K-1}{K}N)$ is shown to be achievable for arbitrary $K$ and $N$ that satisfy $K\ge N$. In this work we will find $K-1$ new points for the case of $N=2$ and $K$ arbitrary. One of these $K-1$ points, namely $(M,R)=(\frac{1}{K},\frac{2(K-1)}{K})$ coincides with the point found in \cite{chen2014fundamental}.
  

For the sake of brevity we skip the formal statement of the problem. It is precisely the setting described in \cite{maddah2014fundamental}. $\;$The rest of the paper is organized as follows: In Section \ref{sec:ex} we will demonstrate the ideas in our caching algorithm via a toy example. In Section \ref{sec:general} we will provide formal description of our placement and delivery strategies for the general case, that is when $K$ is arbitrary and $M = \frac{m}{K}$ for some $m\in\{1,\dots,K-1\}$. In Section \ref{sec:analysis} we prove the correctness of our algorithm and find its achievable rate. Finally, in Section \ref{sec:comparison} we compare the performance of our algorithm with the state of the art techniques introduced in \cite{maddah2014fundamental,chen2014fundamental}.

\section{Example 1}
\label{sec:ex}
We demonstrate our caching strategy via the simplest example which fully represents all the ideas involved in our algorithm. Assume we have $N=2$ files $A$ and $B$, of equal size and $K=6$ users each with a cache of size $M=0.5$ (normalized by file size). We break each file into ${6 \choose 3} = 20 $ parts of equal size and index each part by a set of size three ${\cal T} = \{d_1,d_2,d_3\}$ where $1\le d_1<d_2<d_3\le 6$. User $\ell=1,\dots,6$ stores $A_{\cal T}\oplus  B_{\cal T}$ in its cache if and only if $\ell\in {\cal  T}$. In the following table we represent the content of the cache of each user:
\begin{center}
    \begin{tabular}{| l | l | l | l | l | l | l |}
    \hline
    User 1 & User 2 & User 3 \\ \hline
    $A_{123}\oplus B_{123}$ & $A_{123}\oplus  B_{123}$ & $A_{123}\oplus  B_{123}$\\ 
   $A_{124}\oplus  B_{124}$ & $A_{124}\oplus  B_{124}$  & $A_{134}\oplus  B_{134}$\\ 
   $A_{125}\oplus  B_{125}$ & $A_{125}\oplus  B_{125}$ &  $A_{135}\oplus  B_{135}$\\ 
     $\dots$ & $\dots$  & $\dots$\\ 
      $A_{156}\oplus  B_{156}$ & $A_{256}\oplus  B_{256}$  & $A_{356}\oplus  B_{356}$\\  \hline
      
       \end{tabular}
\end{center}
\begin{center}
    \begin{tabular}{| l | l | l | l | l | l |}
    \hline
   User 4 & User 5& User 6 \\ \hline
    $A_{124}\oplus  B_{124}$ & $A_{125}\oplus  B_{125}$&$A_{126}\oplus  B_{126}$\\ 
 $A_{134}\oplus  B_{134}$ & $A_{135}\oplus  B_{135}$&$A_{136}\oplus  B_{136}$\\ 
$A_{145}\oplus  B_{145}$ & $A_{145}\oplus  B_{145}$&$A_{146}\oplus  B_{146}$\\ 
     $\dots$ & $\dots$  &$\dots$\\ 
$A_{456}\oplus  B_{456}$ & $A_{456}\oplus  B_{456}$&$A_{456}\oplus  B_{456}$\\  \hline
 
       \end{tabular}
\end{center}

Since the size of each subfile is $\frac{1}{20}$ and there are 10 subfiles stored at each cache, the total size of each cache is $M=0.5$. \\
Now assume users $1$ and $2$ ask for file $A$ and users $3,4,5$ and $6$ ask for $B$. In the delivery phase we start by transmitting subfiles of the form $A_{\cal T}$ or $B_{\cal T}$. For each such index ${\cal T}$ we decide whether to transmit $A_{\cal  T}$ or $B_{\cal T}$ depending on how many digits of ${\cal T}$ are from $\{1,2\}$ and how many are from $\{3,4,5,6\}$. More precisely, we fix an integer $0\le j\le KM+1$ and for every such set ${\cal T}$ follow this rule:

If $|{\cal T}\bigcap\{3,4,5,6\}|\ge j$ then transmit $A_{\cal  T}$. Otherwise, transmit $B_{\cal T}$. 

As we will show in Section \ref{sec:analysis}, there is an optimal (not necessarily unique) choice for this parameter $j$ which in our case is $2$. Therefore, we transmit: 
\begin{eqnarray*}
|{\cal T}\bigcap\{3,4,5,6\}|&=& 3:\;\;\; A_{345},A_{346}, A_{356},A_{456}.\\
 |{\cal T}\bigcap\{3,4,5,6\}|&=& 2:\;\;\; A_{134},A_{135}, A_{136},A_{145},\\
&&\hspace{-75.6pt}A_{146},A_{156},A_{234},A_{235}, A_{236},A_{245},A_{246},A_{256}. \\
|{\cal T}\bigcap\{3,4,5,6\}|&=& 1:\;\;\; B_{123},B_{124},B_{125},B_{126}.\\
\end{eqnarray*}
At this stage, each user has access to every subfile he needs except:
\small
\begin{center}
    \begin{tabular}{| l | l | l | l | l | l | l | l | l |}
    \hline
    User 3 & $B_{145}$ & $B_{146}$ & $B_{156}$& $B_{245}$& $B_{246}$&$B_{256}$&$B_{456}$ \\ \hline
    User 4 & $B_{135}$ & $B_{136}$ & $B_{156}$& $B_{235}$& $B_{236}$&$B_{256}$&$B_{356}$\\ \hline
   User 5 & $B_{134}$ & $B_{136}$ & $B_{146}$& $B_{234}$& $B_{236}$&$B_{246}$&$B_{346}$\\ \hline
   User 6 & $B_{134}$ & $B_{135}$ & $B_{145}$& $B_{234}$& $B_{235}$&$B_{245}$&$B_{345}$\\ 
   \hline
      
       \end{tabular}
\end{center}
\normalsize
The last stage of the algorithm is to help each user recover the remaining subfiles. We can transmit (a more formal way of accomplishing this is given below in Section \ref{sec:delivery}).
\begin{eqnarray*}
&&B_{134}\oplus  B_{135}\oplus  B_{145}\;\;\;,\;\;\;  B_{134}\oplus  B_{136}\oplus B_{146}\\
&&B_{135}\oplus  B_{136}\oplus  B_{156}\;\;\;,\;\;\; B_{145}\oplus  B_{146}\oplus  B_{156}\\
&&B_{234}\oplus  B_{235}\oplus  B_{245}\;\;\;,\;\;\;  B_{234}\oplus  B_{236}\oplus  B_{246}\\
&&B_{235}\oplus  B_{236}\oplus  B_{256}\;\;\;,\;\;\;  B_{245}\oplus  B_{246}\oplus  B_{256}\\
&&B_{345}\oplus  B_{346}\oplus  B_{356}\oplus  B_{456}
\end{eqnarray*}
which helps each user in $\{3,4,5,6\}$ recover their desired subfiles. Nevertheless, an important observation here is that the fourth and the eighth messages in the chain above, that is $B_{145}\oplus  B_{146}\oplus  B_{156}$ and $B_{245}\oplus  B_{246}\oplus  B_{256}$  can already be constructed using the earlier transmissions and there is no need to separately transmit them:
\begin{eqnarray*}
&&B_{145}\oplus  B_{146}\oplus  B_{156} = \left(B_{134}\oplus  B_{135}\oplus  B_{145}\right) \oplus\\
&&\left(B_{134}\oplus  B_{136}\oplus  B_{146}\right) \oplus   \left(B_{135}\oplus  B_{136}\oplus  B_{156}\right)
\end{eqnarray*}
and
\begin{eqnarray*}
&&B_{245}\oplus  B_{246}\oplus  B_{256} = \left(B_{234}\oplus  B_{235}\oplus  B_{245}\right) \oplus\\
&&   \left(B_{234}\oplus  B_{236}\oplus  B_{246}\right)\oplus  \left(B_{235}\oplus  B_{236}\oplus  B_{256}\right).
\end{eqnarray*}
Therefore, in total, we are transmitting $27$ messages in the delivery phase which shows we are transmitting at rate $R=\frac{27}{20}$. The worst case achievable rate is obtained by considering all possible choices of different users over $A$ and $B$. In our case, this happens precisely when two users ask for $A$ and the other four ask for $B$ (or vice versa) which is the scenario we studied. This proves that the point $(M,R) = (0.5,\frac{27}{20})$ is achievable when $K=6$ and $N=2$.
\section{Formal Description of the Caching Algorithm}
\label{sec:general}
In this section we describe our caching algorithm for the general case. The setting is as follows: we have $N=2$ files, which we name $A$ and $B$. We have $K$ users each with a cache of size $M = \frac{m}{K}$ for some integer $1\le m \le K-1$. Similar to \cite{maddah2014fundamental} our caching strategy is comprised of two phases: the placement and the delivery phases. We now formally describe each phase.
\subsection{Placement Strategy}
Suppose $M = \frac{m}{K}$ for some integer $1\le m \le K-1$. Partition each file into ${{K}\choose {m}}$ subfiles of equal size and index each subfile with a set of size $m$, i.e. ${\cal T} ={\{d_1,\dots, d_m\}}$ where $1\le d_1< d_2< \dots< d_m\le K$.\\
Store $A_{\cal T}\oplus B_{\cal T}$ at the cache of user $\ell$ if and only if $\ell \in {\cal  T}$. This requires $\frac{{K-1\choose m-1}}{{K \choose m}} = \frac{m}{K}$ bits which is the size of the cache.

\subsection{Delivery Strategy}
\label{sec:delivery}
Without loss of generality, assume the first $L$ users ask for file $A$ and the last $K-L$ users ask for $B$ for some $L\in\{0,\dots,K\}$ (otherwise sort and re-label the users and the subfiles). If $L = K$ or $L = 0$ we transmit all $A_{\cal T}$ or $B_{\cal T}$ subfiles, respectively (therefore, the delivery rate is $R=1$). From here on we assume $L \in \{1,\dots,K-1\}$. The delivery strategy is as follows: Fix an integer $0\le j \le m+1$. Then:

\begin{enumerate}
\item{Transmit  $A_{{\cal T}\bigcup{\cal S}}$  for all sets ${\cal T}$ and ${\cal S}$ such that $|{\cal T}|+|{\cal S}|=m$ and $|{\cal S}|\ge j$ and ${\cal T}\subseteq \{1,\dots,L\}$ and ${\cal S}\subseteq \{L+1,\dots,K\}$.}
\item{Transmit  $B_{{\cal T}\bigcup{\cal S}}$ for all sets ${\cal T}$ and ${\cal S}$ such that $|{\cal T}|+|{\cal S}|=m$ and $|{\cal S}|<j$ and ${\cal T}\subseteq \{1,\dots,L\}$ and ${\cal S}\subseteq \{L+1,\dots,K\}$.}
\item{Transmit ${\cal M}_{{\cal T},{\cal S}} = A_{{\cal T}\bigcup{\cal S}}\oplus \sum_{t\in {\cal T}} A_{ \left(({\cal T}\bigcup\{1\})\backslash\{t\}\right)\bigcup{\cal S}}$ for all sets ${\cal S}$ and ${\cal T}$ such that $|{\cal S}|+|{\cal T}|=m$ and $|{\cal S}|<j$ and ${\cal T}\subseteq \{2,\dots,L\}$ and ${\cal S}\subseteq \{L+1,\dots,K\}$.}
\item{Transmit ${\cal N}_{{\cal T},{\cal S}} = B_{{\cal T}\bigcup{\cal S}}\oplus \sum_{s\in {\cal S}} B_{ {\cal T}\bigcup\left(({\cal S}\bigcup\{L+1\})\backslash\{s\}\right)}$ for all sets ${\cal S}$ and ${\cal T}$ such that $|{\cal S}|+|{\cal T}|=m$ and $|{\cal S}|\ge j$ and ${\cal T}\subseteq \{1,\dots,L\}$ and ${\cal S}\subseteq \{L+2,\dots,K\}$.
}
\end{enumerate}

\section{Analysis}
\label{sec:analysis}
\subsection{Correctness}
We will show that each user is capable of decoding his desired file based on his cache content and based on the messages sent in the delivery phase. Let us concentrate on user $\ell$ for some $\ell \in \{1,\dots,L\}$. The arguments are analogous for $\ell\in \{L+1,\dots,K\}$.\\
Based on the messages sent in step 1 of the delivery phase, user $\ell$ can decode all $A_{{\cal T}\bigcup{\cal S}}$  when $|{\cal S}|\ge j$. From the messages in step 2, user $\ell$ can decode $A_{{\cal T}\bigcup{\cal S}}$ when $|{\cal S}|< j$ and $\ell \in {\cal T}$. What are left to decode after these two phases are $A_{{\cal T}\bigcup{\cal S}}$ when $|{\cal S}|< j$ and $\ell \notin {\cal T}$. If $\ell = 1$, he can decode these messages from ${\cal M}_{{\cal T},{\cal S}} = A_{{\cal T}\bigcup{\cal S}}\oplus \sum_{t\in {\cal T}} A_{ \left(({\cal T}\bigcup\{1\})\backslash\{t\}\right)\bigcup{\cal S}}$ which are sent in step 3 of delivery. If $\ell \neq 1$ but $1\in {\cal T}$, user $\ell$ can again decode $A_{{\cal T}\bigcup{\cal S}}$ from ${\cal M}_{{\cal T}',{\cal S}}$ sent in step 3 of delivery where ${\cal T}' = ({\cal T}\bigcup\{\ell\})\backslash\{1\}$. Assume now that $\ell\neq 1$ and $1\notin{\cal T}$. User $\ell$ forms:

\begin{eqnarray}
&&{\cal M}_{{\cal T},{\cal S}}\oplus \sum_{t\in {\cal T}}M_{({\cal T}\bigcup\{\ell\})\backslash\{t\},{\cal S}}\nonumber\\
&\stackrel{(a)}{=}&A_{{\cal T}\bigcup{\cal S}}\oplus \sum_{t\in {\cal T}}A_{(({\cal T}\bigcup\{\ell\})\backslash\{t\})\bigcup{\cal S}}.
\label{eqn:sumout}
\end{eqnarray}
To establish (a), first note that each term of the form $A_{(({\cal T}\bigcup\{1\})\backslash\{t_1\})\bigcup{\cal S}}$, $t_1\in{\cal T}$ appears exactly twice on the left hand side of the equation, once in ${\cal M}_{{\cal T},{\cal S}}$ and once in $M_{({\cal T}\bigcup\{\ell\})\backslash\{t_1\},{\cal S}}$. Each term of the form $A_{(({\cal T}\bigcup\{1,\ell\})\backslash\{ {t_1,t_2} \})\bigcup{\cal S}}$, $t_1,t_2\in {\cal T}$, $t_1\neq t_2$ also appears exactly twice, once in $M_{({\cal T}\bigcup\{\ell\})\backslash\{t_1\},{\cal S}}$ and once in $M_{({\cal T}\bigcup\{\ell\})\backslash\{t_2\},{\cal S}}$. On the other hand, each term of the form $A_{(({\cal T}\bigcup\{\ell\})\backslash\{t_1\})\bigcup{\cal S}}$, $t_1\in {\cal T}$ appears exactly once in $M_{({\cal T}\bigcup\{\ell\})\backslash\{t_1\},{\cal S}}$. Finally, the term $A_{{\cal T}\bigcup{\cal S}}$ also appears exactly once in ${\cal M}_{{\cal T},{\cal S}}$.
From Equation \eqref{eqn:sumout} user $\ell$ can recover $A_{{\cal T}\bigcup{\cal S}}$ since he knows every other term in the summation. 

\subsection{Achievable Rate}
We count the total number of messages sent in the delivery phase and multiply this by the size of each message that is $\frac{1}{{K \choose m}}$.

First note that each index appears exactly once in the first two steps of delivery. Therefore, the number of messages sent in these two steps is ${K\choose m}$.\\
The number of messages sent in step 3 of delivery is
\begin{eqnarray*}
\#\text{msgs}^3 &=&\sum_{i = 0}^{j-1}  \#\text{msgs}(|S|= i)\\
&=& \sum_{i=\max(0,m-L+1)}^{\min(j-1,K-L)}{K-L\choose i}{L-1\choose m-i}.
\end{eqnarray*}
Similarly, the number of messages sent in step 4 of delivery is
\begin{equation*}
\#\text{msgs}^4 =\sum_{i = \max(j,m-L)}^{\min(m,K-L-1)} {K-L-1 \choose i}{L\choose m-i}.
\end{equation*}
Therefore, the total number of messages sent in the delivery phase multiplied by message size is: 
\begin{eqnarray*}
R_K(M,L,j) &=& 1 + \frac{\sum_{i=\max(0,m-L+1)}^{\min(j-1,K-L)}{K-L\choose i}{L-1\choose m-i}}{{K \choose m}}\\
&+&\frac{\sum_{i =\max(j,m-L)}^{\min(m,K-L-1)} {K-L-1 \choose i}{L\choose m-i}}{{K \choose m}}.
\end{eqnarray*}
We make the following observation: 
\begin{proposition}
There exists a solution to $j^* = \argmin_j R_K(M,L,j)$ that satisfies $j^* = \left\lceil m(1-\frac{L}{K})\right\rceil.$
\end{proposition}
\begin{IEEEproof}
First note that we can restrict $j^*$ to $\max(m-L+1,0)\le j^*\le \min(K-L,m+1)$ since if $j^*<\max(m-L+1,0)$ then $R_K(M,L,j^*)\ge R_K(M,L,\max(m-L+1,0))$ and if $j^*>\min(K-L,m+1)$ then we have $R_K(M,L,j^*) \ge R_K(M,L,\min(K-L,m+1))$.

Next we prove that $j^*\ge m(1-\frac{L}{K})$. If $j^*= \min(K-L,m+1)$ then the inequality is trivial. Assume  $j^*<\min(K-L,m+1)$. Since $j^*$ is optimal, we have $R_K(M,L,j^*+1)\ge R_K(M,L,j^*)$. It follows that:
\small
\begin{eqnarray*}
&&R_K(M,L,j^*+1) - R_K(M,L,j^*) \ge 0\\
&\Rightarrow&{K-L\choose j^*}{L-1 \choose m-j^*}\ge {K-L-1\choose j^*}{L \choose m-j^*}\\
&\Rightarrow&\frac{K-L}{K-L-j^*} - \frac{L}{L-m+j^*}\ge 0\\
&\Rightarrow&j^*\ge m(1-\frac{L}{K}).
\end{eqnarray*}
\normalsize
Finally, we show that $j^*\le m(1-\frac{L}{K})+1$. If $j^*= \max(m-L+1,0)$ then the inequality is trivial. Assume that $j^*> \max(m-L+1,0)$. Then from optimality of $j^*$ it follows that (similar to the previous case) $R_K(M,L,j^*-1)\ge R_K(M,L,j^*)\Rightarrow j^*-1\le m(1-\frac{L}{K})$. Putting these two inequalities together, we obtain $j^* = \left\lceil m(1-\frac{L}{K})\right\rceil.$

\end{IEEEproof}
We can now define 
\begin{eqnarray}
R_K(M,L) &=& 1+\frac{\sum_{i=\max(0,m-L+1)}^{j^*-1}{K-L\choose i}{L-1\choose m-i}}{{K \choose m}}\nonumber\\
&+&\frac{\sum_{i=j^*}^{\min(m,K-L-1)}{K-L-1 \choose i}{L\choose m-i}}{{K \choose m}}
\end{eqnarray}
with $m = MK$ and $j^* = \left\lceil m(1-\frac{L}{K})\right\rceil.$ The achievable rate is the maximum of $R_K(M,L)$ over all possible $1<L<K$:
\begin{equation}
R_K(M) = \max_{0<L<K} R_K(M,L).
\label{eqn:achievablerate}
\end{equation}

\section{Comparison with the State of the Art}
\label{sec:comparison}

\begin{figure}
\includegraphics[scale=0.15]{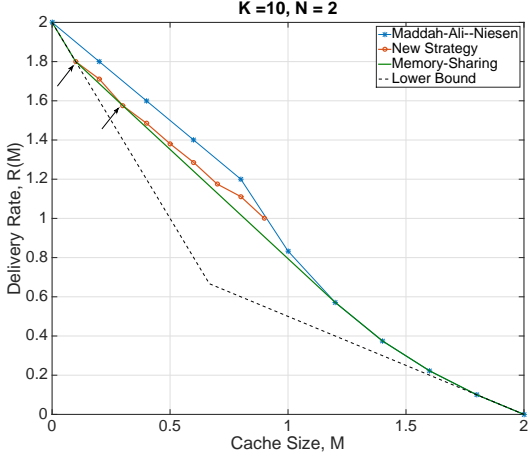}
\caption{Comparison of the achievable rate of our caching strategy with that of Maddah-Ali--Niesen for $K=10$, $N=2$.}
\label{fig:comparison}
\end{figure}

In this section we perform a comparison between the achievable rate of our scheme and that of \cite{maddah2014fundamental,chen2014fundamental}. The achievable rate of our scheme for $K=10$ and $N = 2$ is plotted in red in Figure \ref{fig:comparison} and is found via equation \eqref{eqn:achievablerate} for every $M\in\{\frac{1}{K},\dots,\frac{K-1}{K}\}$. We again emphasize that the leftmost point of our curve, here $(\frac{1}{10},\frac{9}{5})$ has been previously found in \cite{chen2014fundamental}.$\;$ The achievable rate via Maddah-Ali--Niesen caching strategy has been plotted in blue. Evidently from the plot, and as has been proved in the following proposition, our scheme outperforms that of \cite{maddah2014fundamental} for every $M$ at which both rates are defined, that is: $M\in\{\frac{2}{K},\frac{4}{K}\dots,\frac{2(\lceil{\frac{K}{2}\rceil}-1)}{K}\}$ and for every $K$. The proof is in Appendix A. 
\begin{proposition}
Let $R_K(M)$ be our achievable rate as defined in \eqref{eqn:achievablerate}. Let $\hat{R}_K(M)$ be the achievable rate from \cite{maddah2014fundamental}. Then we have:
\small
\begin{eqnarray*}
R_K(M) &\le& \hat{R}_K(M)\;,\; \\
&&\forall K,\; \forall M\in\{\frac{2}{K},\frac{4}{K},\dots,\frac{2(\lceil{\frac{K}{2}\rceil}-1)}{K}\}.
\end{eqnarray*}
\normalsize
The inequality is strict, except when both $K$ is odd and $M = \frac{K-1}{K}$.
\end{proposition}
 For the sake of completeness we have plotted a lower bound on the achievable rate (the dotted black line). Since \cite{maddah2014fundamental}, there has been several works to improve this bound \cite{ghasemi2015improved,sengupta2015improved,tian2015note}. The bound that is plotted here corresponds to the work in \cite{sengupta2015improved} (which in this case coincides with the lower bound given in \cite{maddah2014fundamental}).

\subsection{Memory Sharing and Minimum Size of the Files}

In Figure \ref{fig:comparison} we have also plotted the memory sharing region (green curve). The two points marked by arrows contribute to this region. The first point (found in \cite{chen2014fundamental}) is $(\frac{1}{10},\frac{9}{5})$ and the second point is $(\frac{3}{10},\frac{63}{40})$ The leftmost point from \cite{maddah2014fundamental} that contributes to the memory sharing region is $(\frac{12}{10},\frac{4}{7})$. As $K$ grows large there will be more points $0<M<1$ contributing to the memory-sharing region. For instance when $K = 16$, there are three, namely $(\frac{1}{16},\frac{15}{8})$, $(\frac{3}{16},\frac{97}{56})$, and $(\frac{5}{16},\frac{331}{208})$ and at $K=23$ there are 4 new points. However this dependency is not monotonic in $K$. 

\vspace{2pt}

It is noteworthy that when the file size is small, the other points found by our scheme which lie within the memory sharing region are still relevant. Recall first that our scheme requires each file $A$ and $B$ to be of size at least ${10 \choose m}$ for any particular memory size $M = \frac{m}{K},m\in\{{1},\dots,{K-1}\}$. On the other hand, since the memory sharing strategy interpolates between the points $M = \frac{3}{10}$ and $M = \frac{12}{10}$, the minimum file size for $m\in\{{4},\dots,{9}\}$ must be (see Appendix B)
\small
\begin{eqnarray}
F_{\min}& = &{10 \choose 3}\frac{7(12 - m)}{gcd\{7(12-m),m-3\}}\nonumber\\
&+&{10 \choose 6}\frac{m - 3}{gcd\{7(12-m),m-3\}}
\label{eqn:minbits}
\end{eqnarray}
\normalsize
 which is strictly larger than ${10 \choose m}$ for any $m\in\{{4},\dots,{9}\}$. The difference becomes particularly visible, for instance when $m = {9}$ where memory sharing requires a file size about $100$ times larger than directly applying our caching strategy for $m = 9$.
 
 \section{Conclusion}
The small cache paradigm with much larger number of users than files has not received as much attention in the literature as it deserves. In this work we took a step in improving the achievable rate of this regime by introducing a novel caching strategy for arbitrary number of users and 2 files. Our algorithm takes advantage of simultaneous coded placement and coded delivery to improve upon the achievable rate of \cite{maddah2014fundamental} when the cache is smaller than the size of one file.  Future work will explore the possibility of generalizing our caching algorithm to more than two files.
\section*{Appendix A}

\begin{IEEEproof}[Proof of Proposition 2]

From \cite{maddah2014fundamental}:
\begin{eqnarray*}
\hat{R}_K(M) &=& K(1-\frac{M}{N})\min\{\frac{1}{1+\frac{MK}{N}},\frac{N}{K}\} \\
&=&\begin{cases} 1 &\mbox{ if } K \mbox{ is odd and } M = \frac{K-1}{K},\\ 2 - M &\mbox{ Otherwise. }\end{cases}
\end{eqnarray*}
If $K$ is odd and $M= \frac{K-1}{K}$, then we have
\begin{equation*}
R_K(M,L)= R_K(M,L, j=K-L) =1 = \hat{R}_K(M) . 
\end{equation*}

We will show that if $K$ is even or $M< \frac{K-1}{K}$ then $ R_K(M)< \hat{R}_K(M).$
We consider three cases. First assume $L\ge K-m$. Then we have:
\small
\begin{eqnarray*}
R_K(M,L) &\le& R_K(M,L, j=K-L) \\
&=&1+ \frac{\sum_{i = \max(0,m-L+1)}^{K-L-1}{K-L \choose i}{L-1 \choose m-i}}{{K \choose m}}\\
&<& 1 +  \frac{{K-1\choose m}}{{K \choose m}} = 2 - \frac{m}{K} =  \hat{R}_K(M) . 
\end{eqnarray*}
\normalsize
Next, assume $L < K - m$ and $L< m +1$. Then:
\begin{eqnarray*}
R_K(M,L) &\le& R_K(M,L, j=m - L + 1) \\
&=&1+ \frac{\sum_{i = m-L+1}^{m}{K-L-1 \choose i}{L \choose m-i}}{{K \choose m}}\\
&<& 1 +  \frac{{K-1\choose m}}{{K \choose m}} = 2 - \frac{m}{K} = \hat{R}_K(M). 
\end{eqnarray*}
Finally, assume $L<K - m$ and $L\ge m+1$. Then:
\small
\begin{eqnarray*}
&&R_K(M,L) \le R_K(M,L, j=m)\\
&=&1+ \frac{\sum_{i = 0}^{m-1}{K-L \choose i}{L-1 \choose m-i}+{K-L-1 \choose m}{L \choose 0}}{{K \choose m}}\\
&=& 1 +  \frac{{K-1\choose m}-{K-L \choose m}+{K-L-1 \choose m}}{{K \choose m}}  \\
&<& 2 - \frac{m}{K} = \hat{R}_K(M). 
\end{eqnarray*}
\normalsize
\end{IEEEproof}

\section*{Appendix B}

Proof of Equation \eqref{eqn:minbits}:

First note that the point $M = \frac{3}{10}$ requires ${10\choose 3}$ bits and the points $M = \frac{12}{10}$ needs ${10\choose 6}$ bits (from \cite{maddah2014fundamental}). In order to perform memory sharing for a point $M = \frac{m}{K}$ for some $m\in\{4,\dots,9\}$ we need to first divide file $A$ into two subfiles $A^{(1)}$ and $A^{(2)}$ (same for $B$). We also break the cache into $M^{(1)}$ and $M^{(2)}$ such that $size(M^{(1)}) = \frac{12-m}{m-3}size(M^{(2)})$. Due to the particular caching strategies used at the two end points, we have $size(M^{(1)})=\frac{3}{10}size(A^{(1)})$ and $size(M^{(2)})=\frac{12}{10}size(A^{(2)})$. Therefore,
\begin{eqnarray*}
&&\frac{3}{10} size(A^{(1)})=\frac{12}{10} size(A^{(2)})\frac{12 -m}{m-3}\\
&\Rightarrow&size(A^{(1)}) = \frac{4(12-m)}{m-3}size(A^{(2)}).
\end{eqnarray*}
But $size(A^{(1)})$ must of the form ${10 \choose 3}\ell$ for some integer $\ell$. Similarly,  $size(A^{(2)})={10 \choose 6}\ell'$ for some integer $\ell'$. Thus:
\begin{eqnarray*}
{10 \choose 3}\ell &=& \frac{4(12-m)}{m-3}{10 \choose 6}\ell'\\
&\Rightarrow& \frac{\ell}{\ell'} = \frac{7(12-m)}{m-3}.
\end{eqnarray*}
To choose the smallest $\ell$ and $\ell'$ we have $\ell = \frac{7(12-m)}{gcd\{m-3,7(12-m)\}}$ and $\ell' = \frac{m-3}{gcd\{m-3,7(12-m)\}}$. The claim follows since $F_{\min} = size(A^{(1)}) + size(A^{(2)})$. 

\section*{Acknowledgement}
        \thanks{This work was supported in part by the European ERC Starting Grant 259530-ComCom.}

\bibliographystyle{IEEEtran}
\bibliography{IEEEfull,coded_placement_2}

\end{document}